# Galactic-bursts signatures in Antarctica $^{10}$Be spectra reveal cosmogenesis of climate switching

Mensur Omerbashich, Ph.D.
Physics Department, Faculty of Science, University of Sarajevo
Zmaja od Bosne 33, Sarajevo Bosnia, omerbashich@gmail.com

**Abstract.** A very strong period of 3592±57 yrs in $^{10}$Be deposition rates from Vostok ice-core raw data was detected and verified against concentration raw data at Taylor Dome and Vostok. Data show Hallstadzeit Solar cycle at 2296±57 yrs, and indicate LaViolette period at 12500 yrs. The 99%-confidence Gauss-Vaníček spectral analysis was used, making data alteration avoidable thus enabling data separation that reflected cosmic-ray background conditions at Galactic boundary. After the separation only the new period remains and converges, hence it is of extrasolar and Galactic origin. Since dominant spectral peaks from $^{10}$Be can only be explained by excesses in cosmic-ray influx, the discovered signature indicates bursts occurring regularly in a single source. Based on recent sky surveys, Galactic Core makes the best candidate-host for the bursts. A previously reported 3600-yrs period in geomagnetic field declinations means the discovered phase can overpower astronomical magnetic fields at distances such as Galactic Core – Earth. The epoch of the most recent $^{10}$Be maximum is estimated as 1085±57, coinciding with 1054-1056 alleged account of Crab supernova. The next maximum $^{10}$Be on Earth is predicted in year 4463±57, meaning Earth's climate alternates due to geophysical or non-solar cosmic forcing.

## 1. Introduction

Climatology relies heavily on spectral analyses for modeling and paleostudies. Bard and Frank (2006) give an overview of research in Sun radiation as the extraterrestrial suspect for significant climate changes. But if significant variations in Earth climate are cosmogenic, other sources aside from the Sun could (also) be responsible for such changes. Normally, various proxy records are used for studying past climate. One such proxy is a relatively stable radioactive isotope Beryllium 10 ($^{10}$Be) with a half-life of 1.6 million yrs. It is cosmogenic, i.e., created by irradiation when cosmic rays (carbon, nitrogen, oxygen and other nuclei, i.e., particles charged with very high energy, 100GeV to $10^{15}$ eV) collide with helium and hydrogen ions as the rays travel through Space. Thus $^{10}$Be is believed to not have been created together with most of the Universe matter; this explains why its *in situ* concentrations are minute. Hence, the $^{10}$Be found on Earth is thought to have come mostly from atmospheric fallout, where $^{10}$Be atoms are made by cosmic rays colliding with atmospheric nuclei. The atoms are being carried in snowflakes on to the Earth's surface, where they become part of the ice-sheet via adhering to suspended particles in the water column, or sedimentary records via adsorbing onto available clay particles. So captured, $^{10}$Be atoms are invaluable in climatology where the counting of captured atoms enables detecting the causes of differential influx in cosmic ray radiation; the Sun strength being one such candidate-cause.

Cosmic rays could be responsible for climatic disturbances because cosmic rays make up a significant portion of the energy density of the interstellar medium (Longair, 2002). Many theoretical studies of cosmic rays have been done since pioneering works by Hess, Millikan and Shklovsky. Thus LaViolette (1983, 1987) first proposed that cosmic rays arise from Galactic Center (GC) explosions, suggesting that such bursts can alternate Earth's climate. Observational studies of Galactic TeV γ-ray sources gained an impetus over the last few years with a low-energy-threshold (100 GeV) H.E.S.S. telescope-array (Benbow, 2005, Aharonian et al., 2006, 2005), the INTEGRAL telescope array (Bélanger, 2006) and other telescope GeV/TeV γ-ray surveys of the central 500 parsec region of the GC, as the most unusual part of our Galaxy (Morris and Serabyn, 1996). One of the achievements of ground-based γ-ray astronomy since its inception in 2002, is the recognition of the GC as a TeV γ-ray source (Hofmann, 2005), with more than 30 γ-ray sources detected so far using the H.E.S.S. telescope (Benbow, 2006). One of the most favored models for Galactic sources of high-energy cosmic rays has been the diffusive shock acceleration in supernova remnants (SNR), e.g. Aharonian et al. (2004), but there are serious issues with that model (Hillas 2005, Hofmann 2005).

## 2. Methodology

Climatology faces difficult problems in considering ice-core data such as $^{10}$Be. An important contributor to the error budget is reliability of timescales, with so many different dating methods and consequently many timescales derived in climatology, that Monnin et al. (2004) called into question the whole dating approach. The main problem with methodologies lays in a common approach to data treatment and spectral analyses, where data are prepared so to satisfy limited abilities of the classically used Fourier spectral analysis (FSA) and its derivative methods. Drawbacks of classical approaches such as the signal boosting for very long records, require harsh alterations of data. Researchers often invent data in order to make their analyzed record equispaced; the requirement of the Fourier-based methods. Approaches to data handling and preparation are mostly accommodated to processing algorithms. It is remarkable how easy it is for many researchers to modulate original data, altering many of the unknown yet intrinsic data relationships and distributions. This is one of the most serious issues of all modern science (Omerbashich, 2007a). Its consequence is that researchers rely on ever-more complex algorithms. The data alteration has made observational physical sciences diverge from modeling the nature on to modeling the human errors in understanding the nature, instead.
I regard the use of raw (gapped; unaltered) data as *the* criterion for a physical result's validity (Omerbashich, 2007b), and show that climatology desires such a new viewpoint to data treatment and analysis, already demonstrated as useful (Omerbashich, 2006, 2007a). A natural approach like this helps climatologists who normally work with many polluted or otherwise unreliable data records, and time scales so plentiful yet inconsistent. For example, each time scale sets its own initial stage for inconsistencies in spectral analysis, adding to unreliability of spectra *a priori* affected by drawbacks of the Fourier spectral analysis (FSA) as the most used spectral technique of all science. Unsuitability of FSA is particularly true for long gapped records (Press et al., 2003), such as most records of natural data. As a result of this near disorder, numerous climatology studies range in their conclusions wildly from one extreme ("it's the humans!") to another ("it's Milankovich!").

Due to remoteness and extreme climate the most reliable and consistent records of $^{10}$Be for the past 50 kyrs at least are found in Antarctica ice-cores, Fig. 1. The Vostok $^{10}$Be data are used here as reference data. Vostok is the Antarctica's first scientific station, operating since 1957. The ice sheet thickness at Vostok is around 3700 m. The Byrd and Taylor Dome polar stations, used here for check, both lay within some 2000 km from Vostok. Besides $^{10}$Be ice-core data, various paleoenvironmental datasets with varying timescales and methodologies in data preparation are also being used for verification purposes. Thus I use $^{10}$Be data (deposition-rate; concentration) from the Antarctica stations at Vostok (Raisbeck et al., 1987, 1992) with the LaViolette (1983) timescale against the G4 timescale (Petit et al., 1999, 1997), at Taylor Dome (Steig et al., 1999, 1998) with the accurate st9810 timescale only, at Byrd (Beer et al., 1987) with its original timescale, as well as $^{10}$Be concentration data from 719-2253 m depths at the GISP2 Greenland station (Finkel et al., 1997, Alley et al., 1995, Davis et al., 1990). I used for check the Mg ions dataset from Taylor Dome (Stager et al., 1997, Mayewski et al., 1996, Steig et al., 1999).

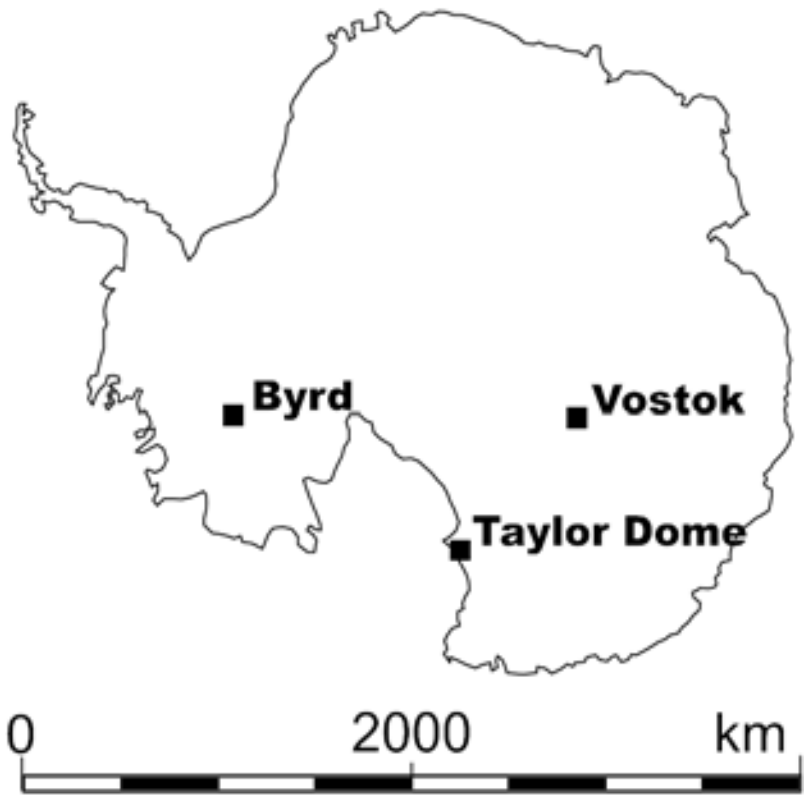

Figure 1. Antarctica: location of world's most reliable $^{10}$Be ice-core data.

Like with all time scales used for same data, the LaViolette (1983) scale, in which relative $^{10}$Be production rate was normalized to the Holocene average value, has a varied reliability. Later on I modify that scale by boosting no peaks in data, contrary to often done modeling of Solar-screening effects during the Holocene period of enhanced Solar activity. Another way of estimating relative cosmic ray intensities outside Solar system is to apply data separation, rather than to edit the data or restricting analyses to a time period when peaks were relatively higher, such as during ice age periods excluding the interglacial. Namely, one can never understand inherently gapped data or portions of such data so well as to be certain that data alterations applied to correct known issues will not introduce a slew of new unknown issues also. Analogously, no other parts of the time series have been corrected for Solar screening, not even during the interstadial intervals, since there is a significant lack of data that forbids applying a model. Situations in which researchers must rely on timescales as described above arise often. In Introduction I argued that a new approach to climatic data analysis is needed: spectral analyses of severely altered yet poorly understood records are dubious at best. Thus given the time-span and density, here only the 99%-confidence level spectral estimates were considered. To get $^{10}$Be deposition rates, $^{10}$Be ice concentration data were corrected for ice accumulation rate tied in with the ice-core assumed chronology (LaViolette, personal communication 2006). No additional corrections were made.

## 3. Results

The strongest detected period in the $^{10}$Be deposition-rate data is 3592±57 yrs, Fig. 2, Table 2 (see Supplementary Data for all Tables). It has not been reported previously, either from ice-core $^{10}$Be analyses, or from tree-ring radiocarbon dating. However, it has been reported at least once in geomagnetic field declinations (Gogorza et al., 1999). The period estimate #3, of 2296±57 yrs, represents the best estimate yet of the *Hallstadzeit Solar cycle*. The Hallstadzeit cycle has been theorized at 2300 yrs, but reported in a range between 2000–2500 yrs, both from ice-core $^{10}$Be studies and in tree-ring dating reports; see e.g. approximate estimates by Tobias et al. (2004), Sonett (1990), and Sonett and Finney (1990). The accuracy of the above estimate is due to using raw instead of reduced data, and to the analysis method able to accommodate such a most natural norm in an optimal way – via least-squares fit. An evidence that the above high-precision estimate is no coincidence, is another extremely high-precision estimate (that of lunar synodic forcing of total-mass Earth) achieved by using same advantages, approach and methodology, from more than ten billion superconducting gravimeter measurements (Omerbashich, 2007b). Clearly, the G-V variance-spectral analysis is the most precise spectral analysis method available, particularly for long gapped records of natural data, or, more generally, for records of data of non-uniform reliability.

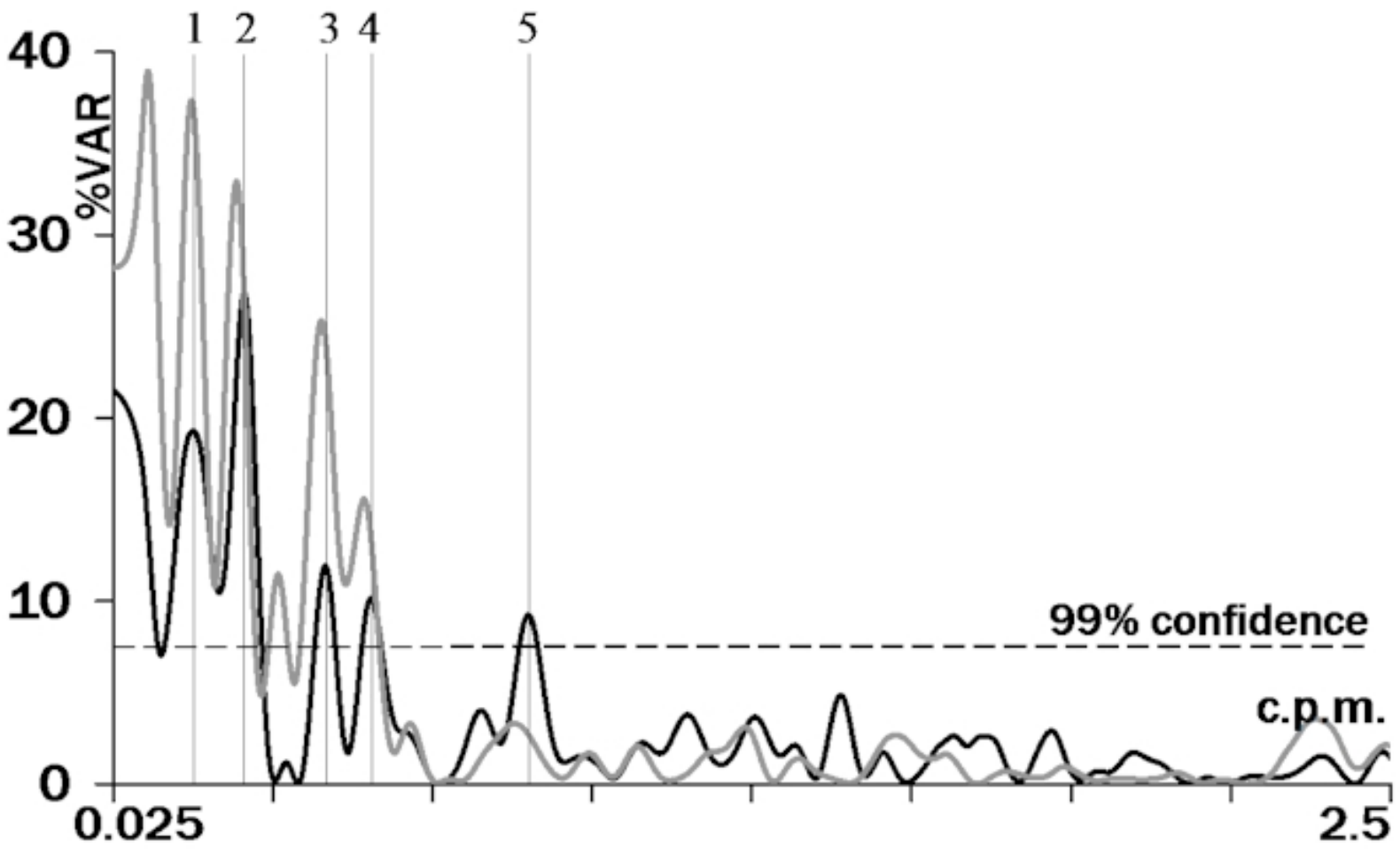

Figure 2. Gauss-Vaníček (G-V) variance-spectrum of the $^{10}$Be Vostok deposition rate raw data with LaViolette (1983) time-scale (black) v. G4 time-scale (gray). Corresponding periods' estimates #1-5 listed in Table 2. Frequency units in cycles per 1 kyr, or cycles per 1 millennium (c.p.m.).

The Hallstadzeit cycle estimate worsened fairly albeit not significantly when the G4 scale is used, Table 3. The period #1, Table 2, has been reported also from a 150-kyrs long Vostok $^{10}$Be record as a coarse "5 kyr" (Liritzis and Gregori, 1997). No periods such as #4 and #5, Table 2, seem to have been reported previously. These may be connected to the multi-century cooling episodes that had occurred every 1500±500 yrs during the Holocene (deMenocal, 2001). The peaks in the G-V spectrum of the Vostok data on the G4 time-scale seem well resolved from both sides, appearing sharp and with underlying background noise suppressed, Fig. 2. Deterioration in estimate of the Hallstadzeit cycle when the G4 timescale is used indicates that G4 timescale creators were more concerned with details, while missing the big picture.

The Vostok $^{10}$Be deposition-rate data show no significant periods at 19000, 40000 or 100000 yrs. LaViolette (1983, 1987) theorized that a ca. 12500 yrs period ought to be the strongest period in $^{10}$Be data. There is a strong background increase in the G-V variance-spectra at around 12500 yrs, Tables 6-9. This period only appears to be poorly resolved, however since the variance-spectrum actually absorbs (feeds on) most of the noise, this strongest and overall persistent cycle represents a verification of the LaViolette period (ibid.) in the strict sense of the Gauss-Vaníček Spectral Analysis (GVSA; see Supplementary Data), and thereby that period's general indication too. In this paper I focus on the overall clearest period found, ca. 3600 yrs.

The $^{10}$Be concentration data reveal no periods longer than ~5500 yrs, and there is no ~3600 yrs period. Empirically, GVSA can discern a period if a data record spans at least four-to-eight times the sought period, under ideal conditions (1:1 noise-signal ratio). However, most of variance had been absorbed by the herein detected and widely spread 40-kyrs period (below 99% confidence), so that this empirical requirement does not necessarily apply. The 40-kyrs period can mean the 39200 yrs obliquity (Berger and Loutre, 1992). Enforcing the 39200 yrs period in the Vostok data however, lowers the strength of all periods considerably; unfortunately, the data span is too short and resolution rather insufficient for the data noise contents to reflect such a long period clearly.

The spectra of cosmic ray proton background radiations (LaViolette, 1983; Fig. 3.7, p.72) show the current electron spectrum and the current proton spectrum, both at Solar minimum. The electron cosmic ray energy density is about 1% of that of the proton cosmic rays, which for protons is $3{\cdot}10^{-2}$ ergs/cm$^2$/s (Ramaty, 1974; Fig. III-15). The other 99% is the cosmic ray proton background, which mostly is of extragalactic origin in that it is isotropically distributed. These are particles emitted from explosive activity that took place in other galaxies. So the marker $^{10}$Be deposition rates that prevailed during the Holocene interval of about 0–10 kyrs BP, and during the previous Eemian interglacial, reflect times when the background was almost entirely extragalactic in origin. Small (weaker) waves might be present, but they are likely lost in the background. Since cosmic ray electrons are much less efficient producers of $^{10}$Be, and since it is believed that there is a significant Solar-screening effect by the heliopause sheath, the cosmic-ray burst wave component might have to rise almost $10^5$-fold before it becomes noticeable above the extragalactic proton background. Subsequently, LaViolette (1983) proposed that a Galactic burst wave exists so energetic as to be able to rise to a peak of $3{\cdot}10^5$ times the current cosmic ray electron background, or to 90 ergs/cm$^2$/sec, and that this wave peak electron intensity would rise to 75 times the current cosmic-ray proton background after taking into account Solar modulation (ibid.; p.233). So during quiescent times, when the burst-wave electron intensity is less than the current cosmic ray proton background, the extragalactic proton background becomes the principle cause of the $^{10}$Be production (cf., ibid.).

Since the use of raw data is regarded herein as *the* criterion for a physical result's validity, one does not correct (alter, basically) data for Solar screening. Instead, data separation was applied at the cutoffs of $2{\cdot}10^5$ atoms/cm$^2$/year for deposition rate, and $0.95{\cdot}10^5$ atoms/gram of ice for concentration dataset, which on the whole should reflect cosmic-ray background conditions at the Galactic boundary during quiescent times (ibid.). Thus the cutoff values were selected such that most of the Holocene values remain below the threshold. At the same time the chosen threshold excludes most of the data from the low cosmic-ray intensity period during the previous interglacial.

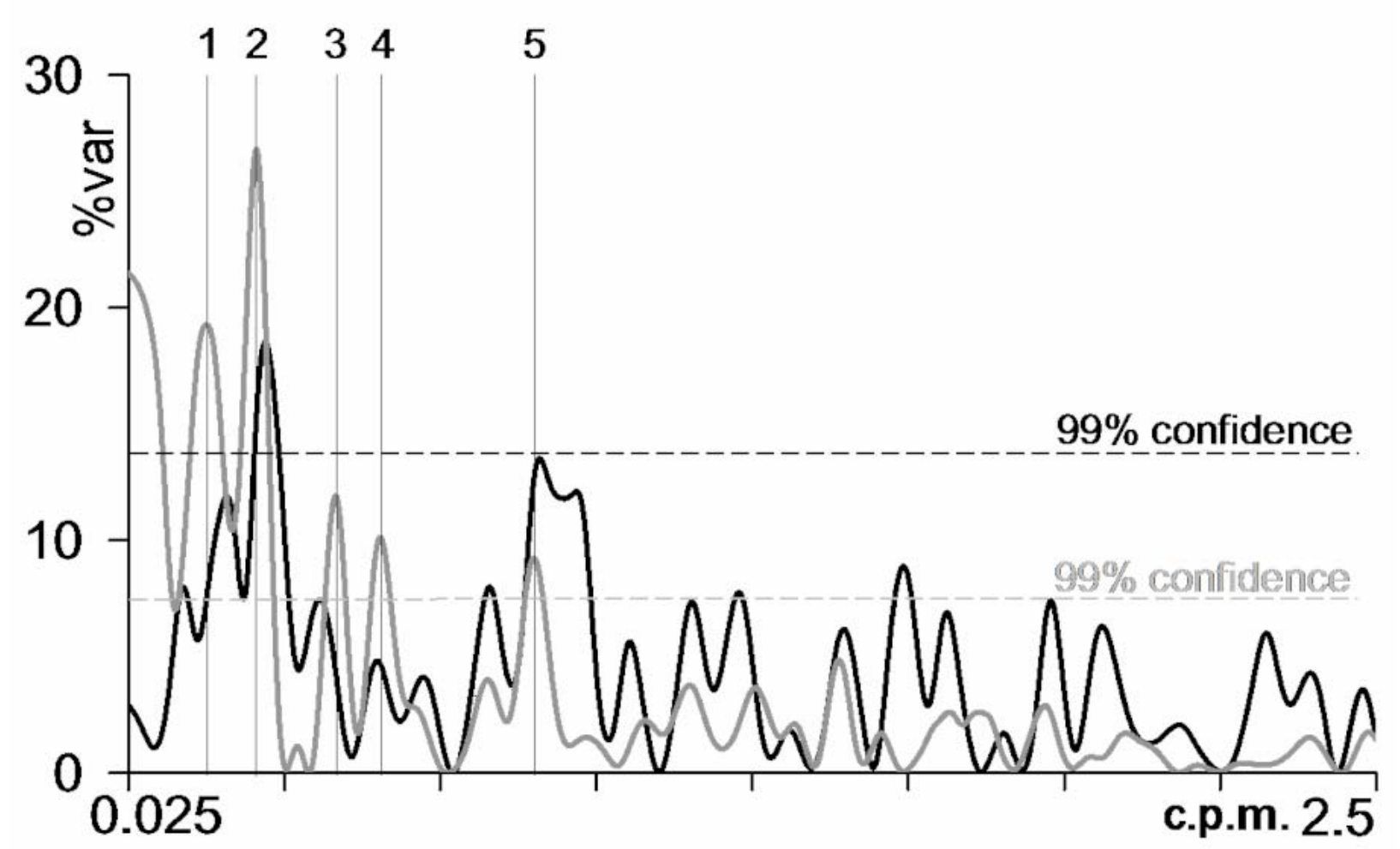

Figure 3. G-V spectra of: Vostok $^{10}$Be deposition rate raw data – all values in Table 1 (gray) v. Vostok $^{10}$Be deposition rate raw at-and-above-cutoff data – boldface values in Table 1 (black). Periods' estimates 1-5 listed in Table 2.

Indeed, the GVSA has revealed, Fig. 3, Tables 7 and 8, that after the separation only the discovered period remains in the at-and-above-cutoff deposition rate and concentration $^{10}$Be data, as: 3378±103 at 19% var, and 3346±85 yrs at 25% var (simple mean value 3362 yrs), respectively. At the same time, the *Hallstadzeit Solar cycle* vanishes in both cases (datasets), which contradicts reports such as by Tobias et al. (2004) who reported the Hallstadzeit cycle in the ice-age portion of the GRIP ice core, probably due to long-periodic noise reflection. Since the Hallstadzeit cycle vanishes after the separation in both the deposition rate and concentration $^{10}$Be data, there is a large sturdiness and purity associated with the 3600 yrs as the strongest period present. Therefore the newly discovered period is genuine. Based on the above, the source of the 3600-yrs period is: (a) extrasolar, because there exists any physically meaningful cutoff value at all, which separates the ca. 3600-yrs and ca. 2300-yrs periods, and (b) Galactic, because the selected cutoff value at the same time reflects background conditions at the Galactic boundary. If the Hallstadzeit cycle does not appear in the ice age portion of a core, this cycle may be more important during periods of Galactic Core quiescence as during the Holocene when the Solar cycle modulates a relatively constant Galactic cosmic ray background, with the exception of the spike around 5300 yrs BP. During the ice age period, or, when the GC becomes more active, extrasolar cosmic-ray activity becomes so much intense as to drown out or overpower this subtle Solar modulation cycle.

I next analyzed only the current quiescent period, from 0–11000 yrs BP, to see if the decoupled 3400-yrs period is absent. If the peak values of a Galactic burst wave for the most part are found below the background, and this period is associated with the burst wave, then that period should not be seen during this more recent interval. The first 11 kyrs of the Vostok $^{10}$Be deposition rate data contains 35 values. GVSA of that portion in 400-5500 yrs band reveals 614 yrs as the only period at 99% and even 95% confidence levels. On the other hand, after the data separation the first 11 kyrs of the at-and-above-cutoff data contains only 5 values. A GVSA of that portion in 400-5500 yrs band reveals no significant periods neither at 99% nor 95% confidence levels. There is a bulge at around 450-650 yrs, but this can be declared a period only based on the above analysis of seven times denser data, not based on statistics for which five values are insufficient. Hence, there appears to be no 3600 (3400) yrs periods in the last 11000 yrs of the Vostok $^{10}$Be.

To finally check for possible effects of the climatic changes, Mg readings from the Taylor Dome core, as a crustal dust indicator, were used, Table 9. Those data should reflect climatic changes which affect the amount of airborne dust, such as wind speed, continental dryness, etc. Note that in addition to periods in Table 9 that are of interest to this study, such as the 3965±16 yrs, many shorter periods were also found with 99% confidence in the 400-114000 yrs band, technically due to the 99% confidence level going all the way down to just 0.3% var. As expected, the Mg records turned out to be far more sensitive to short-term climatic disturbances than the $^{10}$Be records. It can be seen from the analysis of the Mg data that the longest periods have the largest magnitudes, but then also that they are riding on certain unspecified background noise up to 25 var%, which was not enough to burry them beyond significance. This is in agreement with the fact that one should not read magnitude strengths in GVSA literally but in the context of data and physical process those data are unknown to be sensitive to (this is self-evident regardless of spectral analysis method). As already noted, in absolute terms, an outcome in which magnitudes of longest periods from $^{10}$Be concentration data appear twice the magnitudes from $^{10}$Be deposition rate data, means not a contradiction but merely that there are input data issues which are overall unimportant because the longest periods exceeded the 99% confidence level in both cases.

Based on GVSA features and raw-data criterion, contrary to the conventional understanding, timescales represent no real problem for climatology. If so, then most of the error budget in spectral analyses applied to climatology studies comes from a spectral analysis technique one uses. Specifically, the here reported 3600 yrs period is found in all kinds of records ($^{10}$Be, Mg), as well as from various locations (Vostok, Taylor Dome), as well as regardless of the timescale used (except for Byrd, but there were only 73 values there, with the datum not at zero but much later).

## 3. Conclusions

It was showed here that many of differing approaches to timescale definition for core samples analyses not only give inconsistent results, but are also hardly necessary at all: whatever $^{10}$Be dataset or timescale one uses, GVSA produces (or not) more or less the same periods, in most cases. These periods are authentic as they are absolutely strongest and persistent in all records and from various remote locations, even for elements dissimilar to Beryllium, such as Magnesium. Using raw data as a natural approach, remarkable features are revealed in long-periodic (400–40000 yrs) band, from most of $^{10}$Be ice-core datasets: the data are strongly and clearly periodic with a 3600 yrs phase previously unreported from core data, while known periods such as the Hallstadzeit cycle at 2300 yrs were verified with great accuracy also, to the highest degree of accuracy and confidence yet. In addition, a theoretical LaViolette period has been detected for the first time. These two finds give an overall credibility to this analysis as well.

Since the data-separation cutoffs reflected cosmic-ray background conditions at the Galactic boundary, the discovered period represents a signature of cosmic bursts occurring regularly in the GC mid-half kpsc region. A previously reported 3600-yrs period in geomagnetic field declinations strongly supports this conclusion. Using GVSA-specific features, the epoch of the most recent $^{10}$Be maximum on Earth is estimated, as ca. 1085. Given that it takes several decades for $^{10}$Be to accumulate in ice so as to reach detectable levels, this estimated epoch falls remarkably close to the 1054–1056 CE historical account by Asian astronomers of an enormous sky explosion; subsequently, the next maximum level of the $^{10}$Be on Earth is predicted in year 4463±57 CE (see Supplementary Data). Due to relatively long exposure to significant energy rise and the exposure's relatively short recurrence period, it is possible for those recurring Galactic bursts to affect our climate significantly to catastrophically. Then this study indicates that the Earth's significant alternating climate changes arise due to either geophysical or non-Solar cosmic forces.

## Acknowledgments

I thank Dr. Paul LaViolette for his cooperation.

## SUPPLEMENTARY DATA

### Spectral analyses

The raw $^{10}Be$ data were analyzed in the Gauss-Vaníček Spectral Analysis (Vaníček, 1969, 1971) that fits in the least-squares sense data with trigonometric functions. Magnitudes of GVSA spectrum peaks depict the contribution of a period to the variance of a time-series, of the order of some "%" (Vaníček, 1969). As a result, and being based on variance as the most direct measure of noise, the G-V spectra generally depict background noise levels linearly, which gives the GVSA a full physical meaning, making it preferred technique for studying of physical processes and fields simultaneously (Omerbashich, 2007b). In addition, periodicity from incomplete records is estimated down to accuracy prescribed by data, i.e., to the last reliable digit of data values, so that only general reliability (precision) of the input data is superimposed onto a G-V spectrum. At the same time, GVSA precision depends on spectral resolution alone, which however can be chosen at will. Given its character of a general optimizer, least-squares fit will outperform the labeled reliability of data that it fits, most of the time and by as much as 90%. The accompanying statistical analysis in GVSA is intuitive and straightforward, while computer execution-time is a non-issue anymore, with little preprocessing and no post-processing required. The GVSA has been in use for four decades, in various fields such as astronomy, geophysics, medicine, biology, finances, and so on. Taylor and Hamilton (1972) and Omerbashich (2003) have more on the GVSA, including (blind) performance tests. GVSA outperforms Fourier spectral analysis in low-frequency analyses of long records of natural data (Press et al., 2003; Omerbashich, 2006), such as the data used in this study.

A peak in the G-V spectrum is considered well resolved when it is resolved from both sides (slopes), i.e. if the spectrum ascends immediately (in a nearly straight fashion) from zero to the maximum (the spectral peak), and, without retaining its maximum for longer than one spectral point, continuing on to descending in the same manner towards zero. The accuracy of the period's least squares estimate in case of a well-resolved peak can be obtained uniformly for the entire dataset; it equals the declared accuracy of input data, taken over the data density (as: data time-span over number of data input values). Note that any labeled uncertainties in the G-V periods' estimates must not be taken at face-value, as raw data are obviously noise-contaminated. So in reality, and when using records entirely "polluted" (so called in a classical view), the uncertainties could well end up being several times greater than what's labeled, but this is empirical.

As it can be seen from Fig. 2, the G-V period estimates converge rapidly with a magnitude-of-order increase in spectral resolution. Since computations with spectral resolution on the order of $2{\cdot}10^5$ spectral points can put a burden on computing capabilities, subsequently used herein were the $2{\cdot}10^4$ points-resolution throughout the study. Note that the estimate precision stays on the order of (some) years when going from $2{\cdot}10^4$ to $2{\cdot}10^5$ points-resolutions. Note also that the Vostok data, Table 10, for the periods between 0–19 kyrs BP are more densely sampled, with samples taken every 3 to 13 m, whereas after this the sampling interval was at every 25 m for the most part. However, one should not worry about the sample being denser towards its beginning, Table 10, given a relatively small number of values over such a long span anyway. Unless stated otherwise, the band of interest for this analysis is 400–40000 years.

As it can be seen from Tables 2 and 6, the discovered period's estimate gains strength when concentrations are used rather than deposition rates. However, in addition to the peak's magnitude, the underlying long-periodic noise doubles also, so that this magnitude increase in the strongest period found is only apparent. What matters the most is that the periods were found at above 99% confidence level in both cases. As variance naturally measures classical noise, so does % var measures the signal imprint in that noise, and not the signal itself. Whether a signal from one dataset comes out stronger in a GVSA than the same signal from another dataset, where both are at or above 99% confidence level, has little to do with data values themselves but more with the data' overall sensitivity to noise, or the level and the type of noise pollution (reliability issues). Therefore, compared against the Taylor Dome $^{10}Be$ concentration raw data, the representing of $^{10}Be$ raw data as concentrations rather than deposition rates does not significantly affect the accuracy of the period estimate from the GVSA. While the 99% confidence level remained about the same 7.5% var, the signal has doubled (i.e., noise has halved) and the period is now in the 30-40 var% range, while the Taylor signal is in the 20-30 var% range. However, thanks to the GVSA-unique linear representation of background noise (Omerbashich, 2006, 2007b), it can be seen at the same time that the underlying noise too has doubled in magnitude.

It should be expected (see Section 2.) that the Vostok $^{10}Be$ data are at least not less reliable than the Taylor Dome data are; although the periods have doubled, they still remain not any better resolved (qualitative v. quantitative issues of any spectral analysis "art part"). Quite the contrary: there seems to be a background process seen as the degree of unresolved range of periods, in % var, i.e., from 0% var to some % var, that goes up in the same manner as the longest-periodic portion of the band of the drawn periods does. So much so in fact, that it appears to be the representing of some data differently, rather than the accuracy of those data, that affects the signal-to-noise ratio the most. In other words, the doubling of the S-N ratio, as seen in these specific noise-burdened natural records, is truly artificial, since the analysis method used herein does not mind the noise per se – quite contrary. Finally, peaks in a G-V spectrum follow the Boltzmann distribution (Steeves, 1981), so that any change in data representation could perhaps also introduce some Boltzmann-distributed noise as well.

At the same time it appears reasonable to expect that $^{10}Be$ concentration data should reflect not only the 3600-years sensitivity of $^{10}Be$, but of other information contents as well, since a most simple addition of noise effects occurs when using concentrations, meaning the noise-sensitivity adds up uniformly. The deposition rates on the other hand mean by definition the rates of change of $^{10}Be$ only, meaning that effects of the most of other contents arithmetically cancel out. (Note that the two views above do counter a generally accepted view, but so does the herein performed use of noise as the signal; thus the herein posed setup is tautologically correct.) So to represent $^{10}Be$ in terms of deposition rates does act as band-pass filtering, which makes the effect of any concentration magnification get lowered (or halved in this case). Finally, by definition (within this tautological concept), $^{10}Be$ concentration raw data do not reflect changes in the ice accumulation rate. Climatic disorders that suddenly change ice accumulation rates can introduce outliers in data, or can even shift the concentration maximum.

Note that the lowering of the confidence level would not bring much quality to the analysis in our case, as the only new "periods" that emerge in deposition rates at 95% are ~4500-yrs, and one bulge (i.e., not even a glimpsed period) at between 1100-1200 yrs. Note also that 131 deposition rate values were already too small as the number of points (after the separation the number even went down to 71), so that looking at below 99% confidence makes little sense. This was the main reason why 99% was opted for as the minimum confidence level under these circumstances, i.e., given the kind of data, the uncertainty of the timescales, the desired noise pollution, and so on.

In order to estimate the epoch of the most recent increase in the atmospheric $^{10}Be$, and thus extend the classical approach (of simple fit), let us make use of the unique feature of GVSA variance spectra: background noise levels in GVSA depict linearity; this enables relative measurement of field dynamics (Omerbashich, 2007b). It also enables determination of the effect of dataset size on a period estimate. This makes the G-V variance-spectrum an indirect measure of the signal relative strength, as signal-to-noise ratio varies. Basically, one slides a data series for some random small number of years (timestamps permit), say 350 years in our case, then 869, then 1224, then 254, etc. If the period estimate increases at any of these random values, despite the input dataset being shortened, it means that at that moment (datum), the so edited data got fit by the maximum number of cycles possible (maximal S-N ratio), i.e., the so shortened record is virtually precisely at the beginning (datum) of a whole dominant cycle. One repeats this procedure as many times as it makes sense physically, i.e., until the record is shortened so much as to completely extinguish the contribution of the period to variance of the original raw dataset. Then one simply averages all of so obtained estimates of the strongest period in data, which were made each time the period was increasing despite the data actually shortening. Obviously, the higher the data density, the better the results this procedure can give. In our case, this means going more than two full dominant cycles in the past in order to make three matches. Simple-averaging of these three values yielded then an estimate of the last epoch of extremely high $^{10}Be$ on Earth as 1085±57 CE. The results of the epoch estimate are summarized in Table 11. The above-described procedure can be laborious, so selecting datum shifts should be performed in an as random fashion as possible; note however that the likelihood of obtaining any estimate of the last-occurred epoch drops dramatically if systematic datum-shifting is employed instead, Table 11.

Curiously, counting back from year 1085 CE, epoch estimates were obtained for the historical maxima in $^{10}Be$ level, Fig. 7, where the prediction of the next maximum epoch falls in year 4463 CE. We can use the main period's mean value, of 3362 years, obtained after the data separation, Tables 7 & 8. The projected values are seen on Fig. 7 as approximately matching some notable climate altering events, such as those reported by Jessen et al. (2005), as well as related historical events. For instance, the 2227 BCE epoch matches the Akkadian (Mesopotamian) Empire's sudden collapse in the late Holocene around 2220±150 BCE caused by extreme cooling (deMenocal, 2001), as well as the Old Egypt Kingdom's sudden collapse around 2240 BC caused partly by Sahara drying up due to global cooling (Street-Perrott et al., 2000). Next, the 1085 CE epoch matches the Tiwanaku (Bolivian-Peruvian) Empire's collapse at around 1100 CE (deMenocal, 2001). Finally, the 9001 BCE epoch matches the Late Pleistocene extinctions of thousands of species including land-vertebrates such as most of the mammoth genus some 11 000 years ago. Recent radiocarbon dating rejected most of the alternative theories on these extinctions (Guthrie, 2006), which increased plausibility of unnamed cosmic causes to significant climate disturbances on Earth. Other matches may include the herein estimated year 19087 BCE, coinciding with the decline onset of the Wisconsin and Pinedale glacial maximums, around 20 000 years ago (Flannery, 1999). Let it be noted in the end that the herein initially predicted 1085 CE epoch, which matches well with the ancient Asian observation of a remarkable sky event, in terms of historical climatic changes also appears to match the start of the most extreme conditions in the Northern Hemisphere's Medieval Warm Period, around 1000 CE (Cook et al., 2004, Kremenetski et al., 2004, Tiljander et al., 2003).

According to Stephenson and Green (2005), some ten reliable eyewitness accounts have been found in the records covering past two millennia, on what is believed to have had been supernovae explosions. Those records come from ancient Arab, Chinese, Japanese, Korean and North American natives', as well as other eastern and western sources. The most famous of those is a Chinese historical account of a naked-eye observation of an extraordinary sky phenomenon, which had occurred on 4-5 July 1054, and was dubbed "a guest star" by the ancient eyewitnesses. It lasted visibly for some 650 days at the night sky, of which for 23 days in the daylight as well (Sollerman et al., 2001). Some astronomers proposed in the past that the event represented a supernova explosion, the SN1054 remnant of Crab nebula with its Crab pulsar in particular. The Crab nebula is one of the best-studied objects in the sky (*ibid*.). It is not only the brightest optical remnant in the sky, but it is also marked by the most energetic pulsar in the Milky Way. However, according to Peng-Yoke (1962) of Cambridge, western comprehension of ancient Asian astronomical records appears to be quite limited. In addition, Peng-Yoke et al. (1972) have pointed at serious issues about the specific report of 1054, claiming unconditionally that it did not mean an observation of the Crab nebula at all. For instance, (*ibid*.) inspected an ancient Chinese text indicating that a "guest star" had appeared as large as the Moon, which greatly exceeds the normal angular size of any known supernovae. Also, Sollerman et al. (2001) have shown that the Crab supernova did not contain enough mass (Ni) to allow it to burn for nearly two years. Alternative explanations proposed by (*ibid*.), and earlier by Chevalier (1977), would require that, in the sense of mass-energy conversion modeling, the Crab nebula be unique in the entire Universe. This is highly improbable, of course. Then Beilicke et al. (2005) reported of an unidentified extended TeV $\gamma$-ray source named HESS J1303-631, in the Southern Cross region, close to the Galactic plane. This meant the second discovery of a TeV $\gamma$-ray source, following the confirmation on TeV J2032+4130 by Lang et al. (2005), which appears to mean a whole new class of Galactic TeV $\gamma$-ray sources (Aharonian et al., 2005). Finally, H.E.S.S. telescope survey of the innermost 500 parsec of the GC region, by Aharonian et al. (2006), has observed very-high-energy $\gamma$-rays from the GC ridge.

Furthermore, the Galactic supernovae explosions, as described in the historical accounts covering the past millennium, cannot all be accounted for by using what is known on Galactic supernovae rates; see, e.g., Van Den Bergh (1991), and de Donder and Vanbeveren (2003). Since it is highly unlikely that all of the unaccounted-for historical reports meant naked-eye observations of extragalactic supernovae, this leaves room for alternative, Galactic explanations in case of at least some of the above-mentioned historical accounts. However, for lack of alternative explanations of the vast amounts of energy eyewitnessed in 1054, many contemporary astronomers still habitually refer to the 1054 event as the Crab nebula explosion. But beside the optical domain, the energy released from a supernova explosion affects significantly many other energy domains on Earth as well (Iyudin, 2002). Hence, be them authentic or not, the above historical references to supernovae explosions still do set a threshold for this study as well. Specifically, since the energy released from a Galactic supernova explosion evidently does leave terrestrial signatures at atomic scales (*ibid*.), the considerably higher energy emissions from the past GC bursts could have affected the $^{10}$Be records as well, so much so in fact that the period of such a regular event could have become the strongest period in all of the ice-core $^{10}$Be data available, as was the case with the ca. 3600 yrs period discovered here.

**Notes on methodology**

Cosmogenic radionuclides such as $^{10}$Be and $^{14}$C are the most reliable proxies for extending Solar activity reconstructions beyond direct observations of the Sun. High Solar shielding of Galactic cosmic rays during periods of high Solar activity supposedly decreases radionuclide production rates and vice-versa for low Solar activities. The geomagnetic field also influences cosmogenic radionuclide production rates. Similarly to Solar magnetic modulation, high geomagnetic field intensity is thought to decrease the flux of Galactic cosmic rays and radionuclide production rates and vice-versa for low geomagnetic field intensity. The processes responsible for radionuclide production are well known and can be modeled quantitatively. The biggest uncertainties lie in the interpretation of radionuclide records that can be measured in natural archives such as ice cores in the case of $^{10}$Be, or tree rings in the case of $^{14}$C. This because changes in atmospheric transport and deposition in the case of $^{10}$Be, or changes in the carbon cycle in the case of $^{14}$C can influence the measured concentrations.

Unidentified climatic influences lead to errors in reconstructing Solar activity changes using those records. This problem is illustrated by two alternative reconstructions of past changes in Solar activity derived from ice core $^{10}$Be records. Based on a $^{10}$Be record from Antarctica, Bard et al. (2000) conclude that current levels of Solar activity were also reached or exceeded around 1200 CE. By contrast, Solanki et al. (2004) conclude that Solar activity during recent decades is exceptionally high compared to the past 8000 years. A $^{10}$Be record from the Dye 3, Southern Greenland ice core (Beer et al., 1990) seems to corroborate the latter authors' claim. However, the two $^{10}$Be records from Antarctica and Greenland exhibit substantial disagreements for the last 55 years, which seems to be the main reason for such different conclusions (Raisbeck and Yiou, 2004). Obviously at least one of these records must also be influenced by changes in climate. Since $^{14}$C is influenced by completely different geochemical behavior than $^{10}$Be is, investigation of $^{14}$C records can help to solve the contradictions. Since carbon-cycle models allow us to understand past changes in the atmospheric $CO_2$ and $^{13}$C concentrations, it is also possible to use these models to infer the $^{14}$C production rate based on measured $^{14}$C concentrations in tree rings.

For times prior to 1950 CE, when significant amounts of $^{14}$C were released into the atmosphere by nuclear weapons tests, we can calculate variations in the $^{14}$C production rates and infer Solar magnetic modulations from these records. There are uncertainties in connecting the $^{14}$C production rate to recent measurements of Solar magnetic modulation, but regardless of these uncertainties the conclusions by Usoskin et al. (2003) and Solanki et al. (2004) cannot be confirmed by $^{14}$C studies (Muscheler et al., 2005), as $^{14}$C tree ring records indicate that today's Solar activity is high but not exceptional during last 1000 yrs.

Going along a reference timescale used, the coral chronology developed by Bard et al. (2004, 1990) allows for corrections to be made as far back as 20–30 kyrs BP, but there is no definite way to correct the $^{14}$C values for earlier dates as there are no dendrochronology or carved sediment scales available to go by for times beyond about 20 kyrs (generally accepted as a conservative estimate). Dates from 20–70 kyrs BP were based on $^{14}$C as a useful proxy albeit not as reliable as the earlier ice-core dates, since the amount by which $^{14}$C dates should be corrected is unknown. The ice-core dates from 70–14 kyrs BP are based on K/Ar dates, and are thus considered the least reliable in the record. An alternative timescale was checked at the end, which basically was an update of the LaViolette (1983) scale now based on findings by Henderson and Slowey (2000). However, that timescale returned non-converging results using the applied methodology.

The timescale used here as the marker generally gave satisfactory results, such as a one-on-one data separation v. results separation, the dominant period above 99% all the time regardless of the timescale or data density used, the pre-separation v. post-separation difference between two data representations seen as almost an order of magnitude smaller than from the full raw dataset, a whole series of curious matches with significant climate-related historical cataclysmic events, and so on. The same timescale, when updated according to Henderson and Slowey (2000), produced rather distorted results, such as seven "new periods" at 95% confidence level, albeit with a complete mismatch between its 99%-periods in two data representations after the separation, as well as no pre- v. post-separation convergence in the dominant period's estimate from the two ways of data representation, whatsoever.

The $^{10}$Be data from ice cores at Greenland do not show any significant peaks as the raw record is obviously overburdened by long-periodic noise, Fig. 4, perhaps due to a higher snowfall rate than in Antarctica, or significant climate variability on sub-millennium scales. Note however that climate variability in Greenland and Antarctica do not differ significantly on millennium-to-decadal scales (Mayewski et al., 1996). Perhaps first reliable studies on Greenland-Antarctica snowfall and other glacial variability were carried out by ITASE project (Mayewski and Goodwin, 1996). In addition, a southerly-directional influx of extreme cosmic radiation could have further helped to make natural periods remain buried inside the extreme levels of long-periodic noise in the Greenland record. Finally, Beer et al. (1987) did not perform analysis on the Holocene part of the Byrd ice core. Their data begin at around 9200 yrs BP; no sensible results could have been obtained using those data.

## Figures

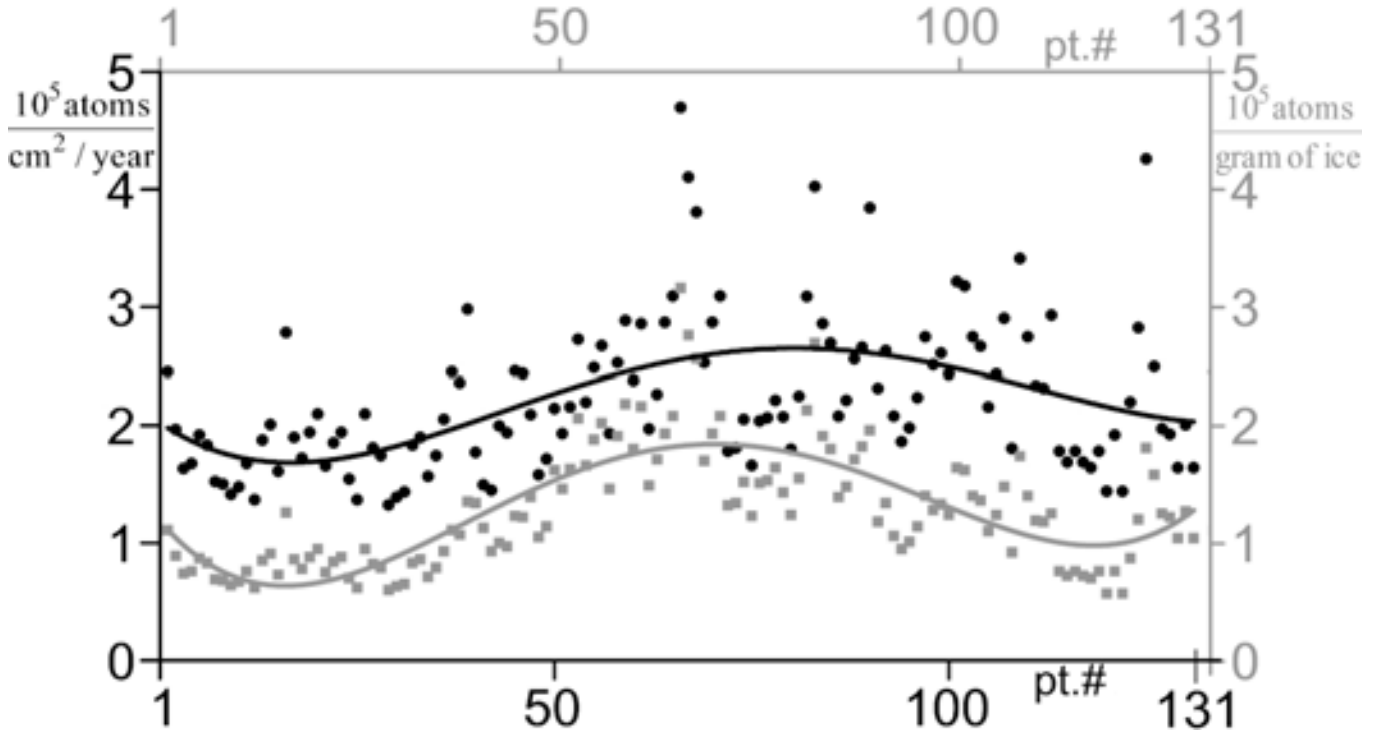

Figure 1. Plot of values, Table 1, of $^{10}$Be data at Vostok (Raisbeck et al., 1987, 1992): deposition rates (black) v. concentrations (gray). To depict the raw data' cyclic tendency, shown is 4$^{th}$ order polynomial fit as a trend over uniformly distributed values (131 dataset points). Note that it is deemed absurd in the realm of the here used raw-data criterion to employ time-plots of raw (gapped!) data.

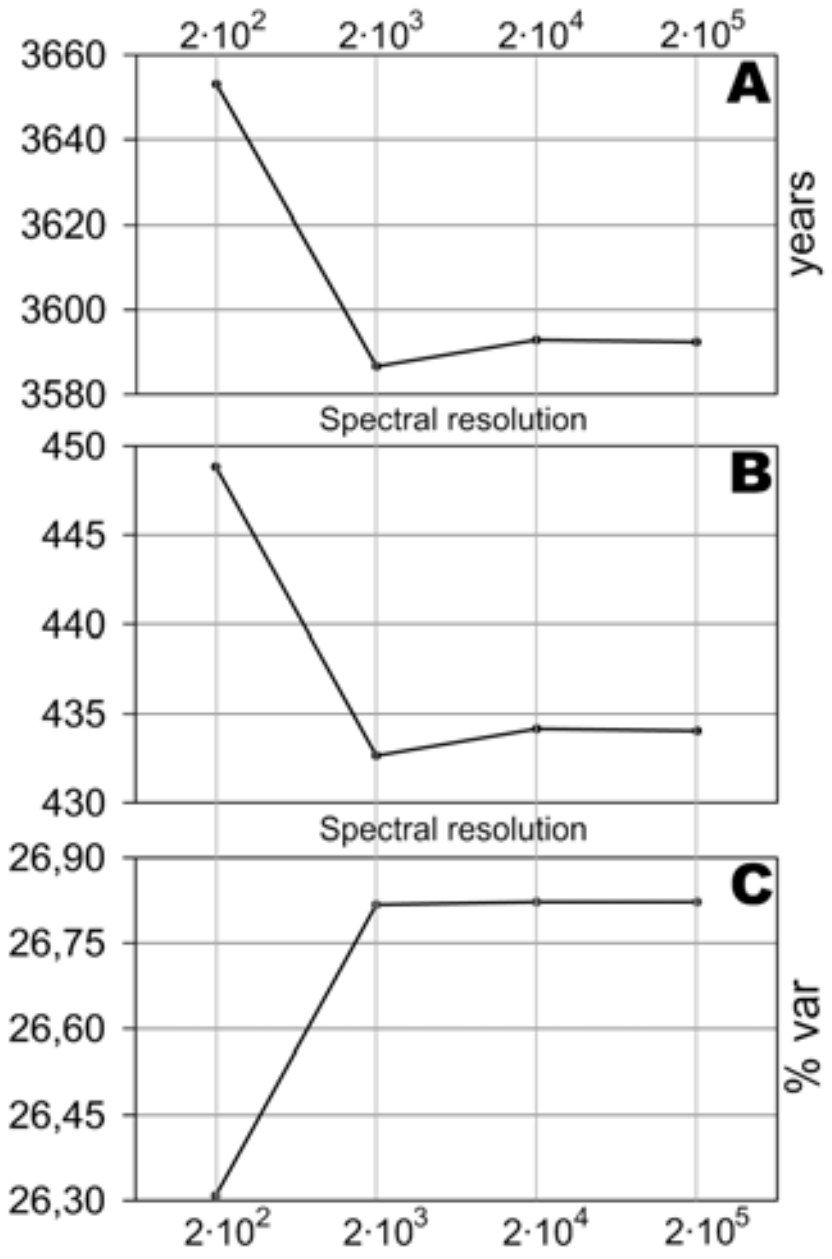

Figure 2. Convergence of GVSA strongest-period estimate with magnitude-of-order increase in spectral resolution. From Vostok $^{10}$Be raw data at 99% confidence level. Spectral resolution in integer number of spectral points (spectral lines). A: period estimate, B: estimate fidelity, C: period magnitude.

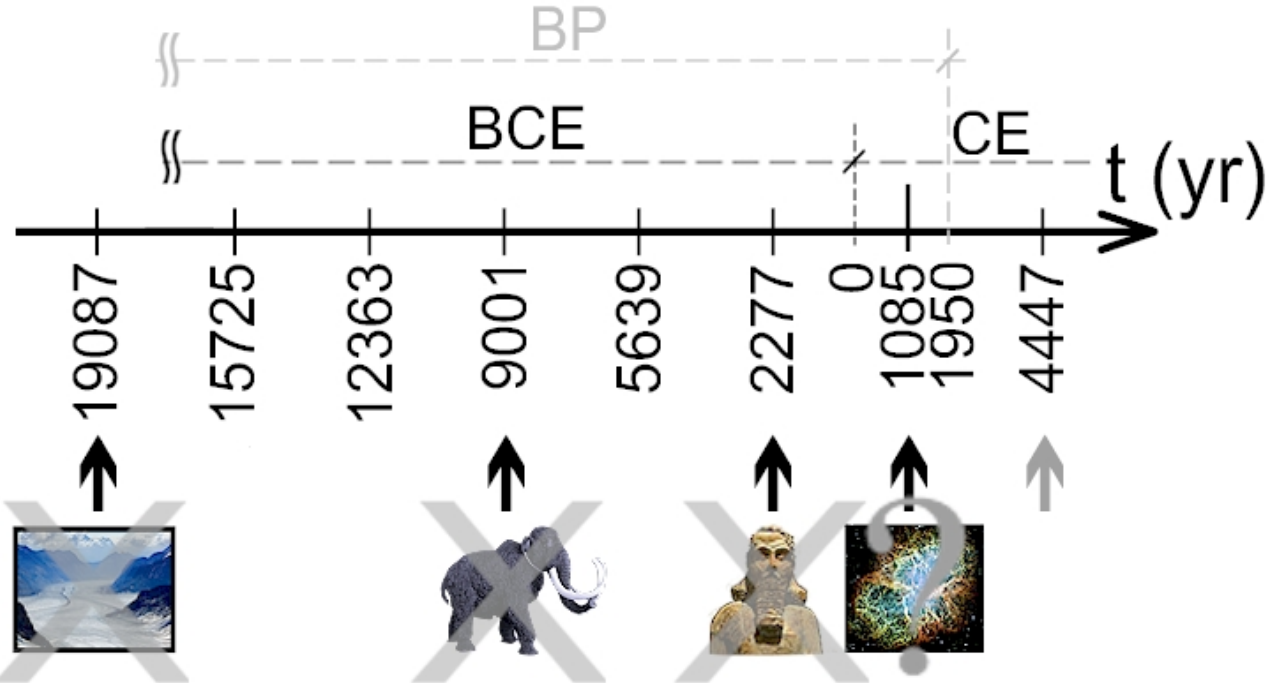

Figure 3. GVSA epoch estimates in years CE/BCE of the maximum levels of $^{10}$Be on Earth. In order to get epochs of the real (atmospheric) maximums, subtract (add to BCE) some decades needed for significant ice accumulation of atmospheric isotope. Arrow markers indicate successful matches of the epoch estimate, Table 1, with known climate-related cultural collapses and extinctions of species (black) and the epoch prediction of the next $^{10}$Be maximum on Earth (gray).

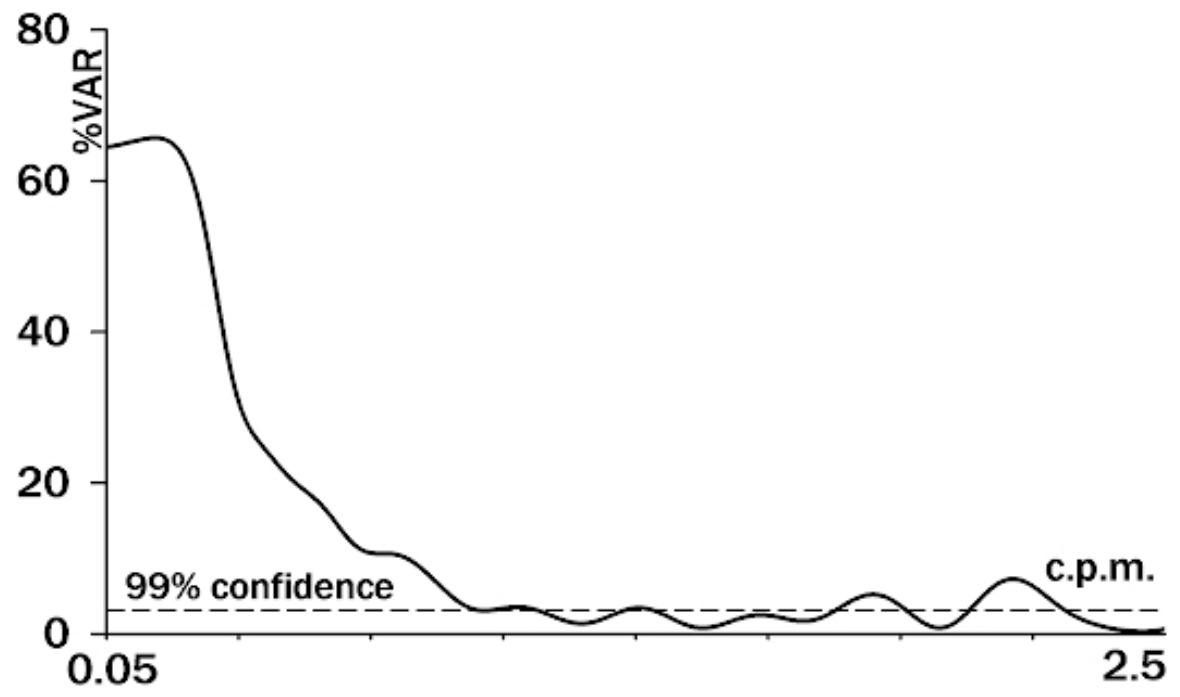

Figure 4. G-V spectrum of GISP2 Greenland record of $^{10}$Be concentration raw data.

**Tables' captions**

Table 1. Vostok $^{10}$Be deposition rate raw data (Raisbeck et al., 1987, 1992), LaViolette (1983) timescale. Values at-and-above $2\cdot10^5$ atoms/cm$^2$/yrs cutoff marked in boldface.

Table 2. GVSA periods found in Vostok $^{10}$Be deposition rate raw data, at 99% confidence level, Fig. 2. Data span: t = 148 663 yrs. Number of input values: N = 131, Table 1, Fig. 2 (Supplement). Data' declared precision: 95%. Period estimate uncertainty (uniform): $\Delta T = (t / N) \cdot 5\% = \pm57$ yrs. Spectral resolution: $2\cdot10^5$ spectral points.

Table 3. GVSA periods found in Vostok $^{10}$Be deposition rate raw data, at 99% confidence level, Fig. 3. Data span: t = 150 972 yrs. Number of input values: N = 131, Table 1, Fig. 2 (Supplement). Data' declared precision: 95%. Period estimate uncertainty (uniform): $\Delta T = (t / N) \cdot 5\% = \pm58$ yrs. Spectral resolution: $2\cdot10^4$ spectral points. G4 timescale.

Table 4. GVSA periods found in Taylor Dome $^{10}$Be concentration raw data, at 99% confidence level. Data span: t = 225 600 yrs. Number of input values: N = 185. Data' declared precision: 95%. Period estimate uncertainty (uniform): $\Delta T = (t / N) \cdot 5\% = \pm61$ yrs. Spectral resolution: $2\cdot10^4$ spectral points. 'st9810' timescale.

Table 5. GVSA periods found in Byrd $^{10}$Be concentration raw data, at 99% confidence level. Data span: t = 32 510 yrs. Number of input values: N = 73. Data' assumed precision: 80%. Period estimate uncertainty (uniform): $\Delta T = (t / N) \cdot 20\% = \pm89$ yrs. Spectral resolution: $2\cdot10^4$ spectral points. Original timescale.

Table 6. GVSA periods found in Vostok $^{10}$Be concentration raw data, at 99% confidence level. Data span: t = 148 663 yrs. Number of input values: N = 131. Data' assumed precision: 95%. Period estimate uncertainty (uniform): $\Delta T = (t / N) \cdot 5\% = \pm57$ yrs. Spectral resolution: $2\cdot10^4$ spectral points. Refers to Tables 7 and 8.

Table 7. GVSA periods found in Vostok $^{10}$Be deposition rate raw data after separation at $2\cdot10^5$ atoms/cm$^2$/yrs, at 95% confidence level. Data span: t = 145 651 yrs. Number of input values: N = 71. Data' assumed precision: 95%. Period estimate uncertainty (uniform): $\Delta T = (t / N) \cdot 5\% = \pm103$ yrs. Spectral resolution: $2\cdot10^4$ spectral points.

Table 8. GVSA periods found in Vostok $^{10}$Be concentration raw data after separation at $0.95\cdot10^5$ atoms/gram of ice, at 95% confidence level. Data span: t = 148 663 yrs. No. of input values: N = 87. Data' assumed precision: 95%. Period estimate uncertainty (uniform): $\Delta T = (t / N) \cdot 5\% = \pm85$ yrs. Spectral resolution: $2\cdot10^4$ spectral points.

Table 9. GVSA periods found in Taylor Dome Mg concentration raw data. Data span: t = 229970 yrs. Number of input values: N = 2969. Data' assumed precision: 80%. Period estimate uncertainty (uniform): $\Delta T = (t / N) \cdot 20\% = \pm16$ yrs. Spectral resolution: $2\cdot10^4$ spectral points.

Table 10. Three estimates of the last epoch when extreme level of $^{10}$Be deposition rates occurred in the Vostok ice-core raw data of $N_0$ values. Spectral resolution 200 000 pt throughout. First $\Delta T_i$ datum: 0.0 yrs BP. Average epoch estimate from 3 locally non-negative (boldface) period changes caused by random datum shifts: 1085±57 CE. This is an example of fairly sufficient shift variability, resulting in three local increases in period estimate despite the reduction in the number of input values.

Table 11. Three estimates of the last epoch when extreme level of $^{10}$Be concentrations occurred in the Taylor Dome ice-core raw data of $N_0$ values. Spectral resolution 200 000 pt throughout. First $\Delta T_i$ datum: 31.5 yrs BP. Average epoch estimate from locally non-negative period changes as caused by random datum shifts not possible. This is an example of insufficient datum-shift variability, resulting in no local increase in period estimate (with reducing the number of input data values).

**Tables**

| Value # | $t_i$ [years BP] | $^{10}$Be $\left[\frac{\frac{\text{atoms}}{10^{-5}}}{\frac{\text{cm}^2}{\text{year}}}\right]$ | Value # | $t_i$ [years BP] | $^{10}$Be $\left[\frac{\frac{\text{atoms}}{10^{-5}}}{\frac{\text{cm}^2}{\text{year}}}\right]$ | Value # | $t_i$ [years BP] | $^{10}$Be $\left[\frac{\frac{\text{atoms}}{10^{-5}}}{\frac{\text{cm}^2}{\text{year}}}\right]$ |
|---|---|---|---|---|---|---|---|---|
| 001 | 00.00 | **2.45** | 045 | 14.02 | **2.46** | 089 | 67.92 | **2.66** |
| 002 | 00.43 | 1.96 | 046 | 14.39 | **2.44** | 090 | 70.00 | **3.84** |
| 003 | 01.12 | 1.63 | 047 | 15.23 | **2.09** | 091 | 71.75 | **2.31** |
| 004 | 01.26 | 1.68 | 048 | 15.49 | 1.58 | 092 | 73.50 | **2.63** |
| 005 | 01.50 | 1.92 | 049 | 15.52 | 1.72 | 093 | 75.24 | **2.08** |
| 006 | 02.30 | 1.83 | 050 | 15.85 | **2.14** | 094 | 76.99 | 1.86 |
| 007 | 02.57 | 1.52 | 051 | 16.67 | 1.93 | 095 | 78.74 | 1.98 |
| 008 | 02.92 | 1.50 | 052 | 17.34 | **2.16** | 096 | 80.49 | **2.24** |
| 009 | 02.93 | 1.41 | 053 | 17.53 | **2.72** | 097 | 82.24 | **2.75** |
| 010 | 03.35 | 1.48 | 054 | 18.36 | **2.20** | 098 | 83.99 | **2.51** |
| 011 | 04.16 | 1.68 | 055 | 18.58 | **2.49** | 099 | 85.73 | **2.61** |
| 012 | 04.41 | 1.37 | 056 | 19.23 | **2.67** | 100 | 87.48 | **2.43** |
| 013 | 04.42 | 1.88 | 057 | 19.49 | 1.93 | 101 | 89.23 | **3.22** |
| 014 | 04.72 | **2.01** | 058 | 21.13 | **2.53** | 102 | 90.98 | **3.18** |
| 015 | 05.06 | 1.61 | 059 | 23.02 | **2.88** | 103 | 92.73 | **2.75** |
| 016 | 05.34 | **2.78** | 060 | 24.92 | **2.38** | 104 | 94.48 | **2.67** |
| 017 | 05.73 | 1.90 | 061 | 26.81 | **2.86** | 105 | 96.22 | **2.16** |
| 018 | 05.79 | 1.72 | 062 | 28.70 | 1.97 | 106 | 97.97 | **2.43** |
| 019 | 06.03 | 1.94 | 063 | 30.60 | **2.26** | 107 | 99.72 | **2.90** |
| 020 | 06.28 | **2.10** | 064 | 32.35 | **2.87** | 108 | 101.47 | 1.80 |
| 021 | 06.76 | 1.66 | 065 | 33.55 | **3.09** | 109 | 103.22 | **3.41** |
| 022 | 06.86 | 1.85 | 066 | 34.11 | **4.70** | 110 | 104.97 | **2.75** |
| 023 | 07.27 | 1.94 | 067 | 34.68 | **4.10** | 111 | 106.71 | **2.33** |
| 024 | 07.61 | 1.54 | 068 | 35.66 | **3.81** | 112 | 108.46 | **2.31** |
| 025 | 07.79 | 1.37 | 069 | 35.94 | **2.53** | 113 | 110.35 | **2.93** |
| 026 | 07.94 | **2.10** | 070 | 36.58 | **2.87** | 114 | 112.08 | 1.78 |
| 027 | 08.49 | 1.81 | 071 | 37.63 | **3.09** | 115 | 113.82 | 1.69 |
| 028 | 08.69 | 1.74 | 072 | 39.64 | 1.78 | 116 | 115.56 | 1.78 |
| 029 | 09.01 | 1.32 | 073 | 41.69 | 1.81 | 117 | 117.29 | 1.69 |
| 030 | 09.20 | 1.39 | 074 | 43.74 | **2.05** | 118 | 119.03 | 1.64 |
| 031 | 09.69 | 1.43 | 075 | 45.79 | 1.66 | 119 | 120.76 | 1.78 |
| 032 | 09.96 | 1.83 | 076 | 47.84 | **2.04** | 120 | 122.50 | 1.44 |
| 033 | 10.08 | 1.90 | 077 | 49.89 | **2.07** | 121 | 124.22 | 1.92 |
| 034 | 10.49 | 1.57 | 078 | 51.93 | **2.21** | 122 | 125.95 | 1.44 |
| 035 | 10.92 | 1.74 | 079 | 53.98 | **2.07** | 123 | 127.67 | **2.20** |
| 036 | 11.16 | **2.05** | 080 | 55.98 | 1.80 | 124 | 129.54 | **2.82** |
| 037 | 11.21 | **2.45** | 081 | 57.97 | **2.25** | 125 | 131.46 | **4.26** |
| 038 | 11.35 | **2.36** | 082 | 59.16 | **3.09** | 126 | 133.60 | **2.49** |
| 039 | 11.43 | **2.98** | 083 | 59.95 | **4.02** | 127 | 136.61 | 1.97 |
| 040 | 12.30 | 1.77 | 084 | 60.75 | **2.86** | 128 | 139.63 | 1.92 |
| 041 | 12.70 | 1.49 | 085 | 61.14 | **2.69** | 129 | 142.64 | 1.64 |
| 042 | 13.15 | 1.45 | 086 | 61.94 | **2.08** | 130 | 145.65 | **2.00** |
| 043 | 13.37 | **2.00** | 087 | 63.92 | **2.21** | 131 | 148.66 | 1.64 |
| 044 | 13.95 | 1.94 | 088 | 65.90 | **2.56** | 131 | 148.66 | 1.64 |

Table 1

Table 2

| # | period [yr] | fidelity | mag (var%) | power (dB) |
|---|---|---|---|---|
| 1 | 5583 | 1048 | 19.3 | -6.2 |
| 2 | 3592 | 434 | 26.8 | -4.4 |
| 3 | 2296 | 177 | 11.9 | -8.7 |
| 4 | 1908 | 122 | 10.1 | -9.5 |
| 5 | 1206 | 49 | 9.2 | -9.9 |

Table 3

| # | period [yr] | fidelity | mag (var%) | power (dB) |
|---|---|---|---|---|
| 1 | 12427 | 5115 | 38.9 | -1.9 |
| 2 | 6060 | 1216 | 37.4 | -2.2 |
| 3 | 3950 | 517 | 32.9 | -3.1 |
| 4 | 2993 | 297 | 11.5 | -8.9 |
| 5 | 2390 | 189 | 25.3 | -4.7 |
| 6 | 1995 | 132 | 15.6 | -7.3 |

Table 4

| # | period [yr] | fidelity | mag (var%) | power (dB) |
|---|---|---|---|---|
| 1 | 18448 | 7544 | 25.9 | -4.6 |
| 2 | 10846 | 2608 | 28.2 | -4.1 |
| 3 | 4305 | 411 | 21.5 | -5.6 |
| 4 | 3825 | 324 | 21.5 | -5.6 |
| 5 | 2251 | 112 | 28.7 | -3.9 |
| 6 | 1701 | 64 | 5.6 | -12.2 |

Table 5

| # | period [yr] | fidelity | mag (var%) | power (dB) |
|---|---|---|---|---|
| 1 | 5503 | 4657 | 28.7 | -3.9 |
| 2 | 1570 | 379 | 46.8 | -0.6 |
| 3 | 950 | 139 | 19.5 | -6.2 |

Table 6

| # | period [yr] | fidelity | mag (var%) | power (dB) |
|---|---|---|---|---|
| 1 | 12563 | 5304 | 44.4 | -1.0 |
| 2 | 5926 | 1181 | 30.3 | -3.6 |
| 3 | 3721 | 466 | 38.7 | -2.0 |
| 4 | 2291 | 176 | 18.1 | -6.6 |
| 5 | 1935 | 126 | 12.5 | -8.5 |

Table 7

| # | period [yr] | fidelity | mag (var%) | power (dB) |
|---|---|---|---|---|
| 1 | 3378 | 392 | 18.5 | -6.4 |

| # | period [yr] | fidelity | mag (var%) | power (dB) |
|---|---|---|---|---|
| 1 | 13246 | 5902 | 22.3 | -5.4 |
| 2 | 7953 | 2127 | 23.8 | -5.1 |
| 3 | 4428 | 659 | 16.7 | -7.0 |
| 4 | 3346 | 376 | 25.2 | -4.7 |
| 5 | 1651 | 92 | 11.8 | -8.8 |
| 6 | 1359 | 62 | 14.3 | -7.8 |
| 7 | 831 | 23 | 15.8 | -7.3 |

Table 8

| # | period [yr] | fidelity | mag (var%) | power (dB) |
|---|---|---|---|---|
| 1 | 11694 | 2973 | 39.3 | -1.9 |
| 2 | 3965 | 342 | 43.6 | -1.1 |
| 3 | 2464 | 132 | 35.0 | -2.7 |
| 4 | 1698 | 63 | 11.6 | -8.8 |

Table 9

| shift # | $T_i$ @99%conf [yrs] | $\Delta T_i = T_{i+1} - T_i$ [yrs] | $mag_i$ [var%] | fidelity | $N_i$ | $(N_o - N_i) < (N_o \times mag_o)$ | 1/dens= span/N [yrs] | datum $shift_i$ [yrs] | $b_i = shift_i - shift_{i+1}$ [yrs BP] | epoch = $b_i$+ 1950 [yrs CE] |
|---|---|---|---|---|---|---|---|---|---|---|
| **0 (raw)** | **(3592.7)** | **+0.9** | **(26.8)** | **(434.0)** | **131** | **yes** | **(1135)** | **(427.0)** | **-697.4** | **1253** |
| 1 | 3593.6 | -5.0 | 26.8 | 435.6 | 130 | yes | 1140 | 1124.5 | -375.5 | |
| 2 | 3588.6 | -17.0 | 26.9 | 436.4 | 129 | yes | 1144 | 1500.0 | -796.1 | |
| 3 | 3571.6 | -5.2 | 27.6 | 433.4 | 127 | yes | 1159 | 2296.1 | -622.8 | |
| 4 | 3566.4 | -15.4 | 27.7 | 434.5 | 126 | yes | 1162 | 2918.9 | -430.9 | |
| 5 | 3550.9 | -22.3 | 28.1 | 432.6 | 124 | yes | 1175 | 3349.8 | -811.2 | |
| 6 | 3528.6 | -11.1 | 28.8 | 428.4 | 122 | yes | 1191 | 4160.9 | -557.9 | |
| 7 | 3517.5 | -24.6 | 29.1 | 428.1 | 121 | yes | 1194 | 4718.9 | -620.6 | |
| 8 | 3493.0 | -12.1 | 29.6 | 423.8 | 118 | yes | 1220 | 5339.5 | -385.8 | |
| **9** | **3480.8** | **+8.9** | **29.7** | **422.7** | **116** | **yes** | **1236** | **5725.3** | **-1030.0** | **920** |
| 10 | 3489.8 | -14.8 | 30.0 | 426.0 | 115 | yes | 1243 | 6755.4 | -512.4 | |
| 11 | 3475.0 | -11.8 | 29.8 | 425.5 | 111 | yes | 1278 | 7267.8 | -346.8 | |
| 12 | 3463.2 | -2.5 | 29.5 | 424.1 | 109 | yes | 1297 | 7614.6 | -323.2 | |
| 13 | 3460.6 | -22.2 | 29.3 | 424.5 | 108 | yes | 1306 | 7937.8 | -556.2 | |
| 14 | 3438.4 | -0.3 | 28.9 | 420.1 | 106 | yes | 1328 | 8494.0 | -516.7 | |
| 15 | 3438.1 | -11.1 | 28.8 | 421.7 | 105 | yes | 1335 | 9010.7 | -682.0 | |
| 16 | 3427.0 | -29.5 | 28.2 | 420.5 | 103 | yes | 1356 | 9692.7 | -391.0 | |
| 17 | 3397.5 | -20.5 | 27.6 | 415.3 | 101 | yes | 1376 | 10083.7 | -401.7 | |
| 18 | 3377.0 | -5.4 | 26.9 | 411.4 | 99 | yes | 1400 | 10485.4 | -431.3 | |
| 19 | 3371.6 | -13.1 | 26.6 | 411.3 | 98 | yes | 1410 | 10916.7 | -512.9 | |
| **20** | **3358.5** | **+4.3** | **26.2** | **409.4** | **97** | **yes** | **1420** | **11429.6** | **-869.3** | **1081** |
| 21 | 3362.8 | | | | 93 | no (STOP) | | | | |
| **Average epoch CE: 1085±57** | | | | | | | | | | |

Table 10

| shift # | $T_i$ @99%conf [yrs] | $\Delta T_i = T_{i+1} - T_i$ [yrs] | $mag_i$ [var%] | fidelity | $N_i$ | $(N_o - N_i) < (N_o \times mag_o)$ | 1/dens= span/N [yrs] | datum $shift_i$ [yrs] | $b_i = shift_i - shift_{i+1}$ [yrs BP] | epoch = $b_i$+ 1950 [yrs CE] |
|---|---|---|---|---|---|---|---|---|---|---|
| 0 (raw) | 3825.7 | -14.5 | 21.5 | 324,4 | 185 | yes | (1219) | 341.1 | -310.2 | |
| 1 | 3811.2 | -13.3 | 20.8 | 322.5 | 179 | yes | 1258 | 651.3 | -266.0 | |
| 2 | 3797.9 | -2.0 | 20.3 | 320.7 | 175 | yes | 1285 | 917.3 | -291.0 | |
| 3 | 3795.9 | -11.1 | 20.0 | 320.7 | 172 | yes | 1306 | 1208.3 | -307.7 | |
| 4 | 3784.7 | -20.4 | 19.6 | 319.2 | 169 | yes | 1328 | 1516.0 | -351.9 | |
| 5 | 3764.5 | -23.1 | 19.1 | 316.3 | 166 | yes | 1350 | 1867.9 | -325.5 | |
| 6 | 3741.4 | -14.4 | 18.7 | 312.9 | 163 | yes | 1372 | 2193.4 | -365.9 | |
| 7 | 3727.0 | -10.0 | 18.2 | 310.9 | 161 | yes | 1387 | 2559.3 | -257.8 | |
| 8 | 3717.0 | -2.8 | 17.6 | 309.8 | 158 | yes | 1411 | 2817.1 | -259.2 | |
| 9 | 3714.2 | -3.9 | 17.3 | 309.7 | 157 | yes | 1419 | 3076.3 | -247.1 | |
| 10 | 3710.3 | -3.2 | 16.8 | 309.4 | 155 | yes | 1435 | 3323.4 | -310.9 | |
| 11 | 3707.1 | -2.6 | 16.2 | 309.2 | 153 | yes | 1453 | 3634.3 | -300.1 | |
| 12 | 3704.5 | -2.7 | 15.7 | 309.2 | 151 | yes | 1470 | 3934.4 | -309.6 | |
| 13 | 3701.8 | -1.5 | 14.7 | 309.1 | 148 | yes | 1498 | 4244.0 | -287.1 | |
| 14 | 3700.2 | -0.7 | 14.1 | 309.3 | 146 | yes | 1516 | 4531.1 | -386.1 | |
| 15 | 3699.6 | | | | 144 | no (STOP) | | | | |
| **Average epoch CE:** | | | **N/A** | | | | | | | |

Table 11